\documentclass[aps,prb,twocolumn,showpacs,preprintnumbers]{revtex4}
\usepackage{graphicx}
\usepackage{bm}
\usepackage{color}
\usepackage{amsmath}
\usepackage{epsfig}
\usepackage{epstopdf}
\usepackage{comment}

\begin{document}
\title{ 
Interplay between Spin-Orbit Coupling and Strong Correlation Effects:\\
Comparison of Three Osmate Double Perovskites: Ba$_{2}A$OsO$_{6}$ ($A=$Na, Ca, Y)}
\author{Shruba Gangopadhyay and Warren E. Pickett }
\affiliation{Department of Physics, University of California, Davis CA 95616, USA}

\begin{abstract}
High formal valence Os-based double perovskites are a focus of current 
interest because they display strong interplay of large spin orbit coupling 
and strong electronic correlation. Here we present the electronic and magnetic 
characteristics of a sequence of three cubic Os based double perovskites Ba$_{2}$$A$OsO$_{6}$, 
($A=$Na, Ca, Y), with formal valences (charge states) of Os$^{+7}~(d^{1})$, 
Os$^{+6}(d^{2})$, Os$^{+5}(d^{3})$. 
For these first principles based calculations 
we apply an ``exact exchange for correlated electrons'' functional, with exact exchange
applied in a hybrid fashion solely to the Os($5d$) states. While Ba$_{2}$NaOsO$_{6}$ is a 
ferromagnetic Dirac-Mott insulator studied previously, the other two show 
antiferromagnetic ordering while all retain the cubic (undistorted) structure. 
For comparison purposes we have investigated only the ferromagnetic ordered 
phase. 
A metal-insulator transition is predicted
to occur upon rotating the direction of magnetization in all three materials, reflecting 
the central role of spin orbit coupling in these small gap osmates. 
Surprises arising from comparing formal charge states with the radial 
charge densities are discussed. Chemical shielding factors and 
orbital susceptibilities are provided for comparison with future nuclear magnetic resonance data. 
\end{abstract}
\date{\today}
\maketitle

\section{Introduction}

Strong spin-orbit coupling (SOC) is a relativistic effect that couples the orbital angular momentum to the electron spin in atoms, and can often be treated as a small perturbation in the discussion of electrons in solids. However, in heavy elements SOC is not weak, and indeed can lead to striking qualitative effects. 
Recent interest is largely in the diverse properties of electronic materials that are insulating, or in the process of becoming so, as a result of electron-electron interactions, the most prominent one being the strong local Hubbard repulsion $U$ between electrons within an open subshell. For these heavy ions SOC may becomes a competing or even dominating factor, mixing the various spin, orbital, charge, and lattice degrees of freedom. 

The interplay of  strong electron correlation and large SOC is relatively less explored, and certainly not well understood  at all, because the behavior involves so many comparable energy scales. This situation arises in a broad  family of magnetic Mott insulating systems in which three-fold degenerate $t_{2g}$ orbitals  are partially filled.\cite{balentsprb1} In such systems orbital degeneracy is protected only by cubic lattice symmetry, and typically the crystal field splitting is large enough that $e_{g}$ orbitals are out of the picture. Insulating states sometimes arise due to spin-orbit coupling, as 
shown\cite{bnoo_prb} for Ba$_2$NaOsO$_6$. This importance of SOC classified BNOO as a Dirac-Mott insulator.\cite{balents_dm} Upon descending the periodic table from the 3\textit{d} to 4\textit{d} to the 5\textit{d} series, there are several competing trends. First, the \textit{d} orbitals become more extended, tending to reduce the electronic repulsion $U$ and thereby diminish correlation effects.
However, simultaneously, the SOC increases dramatically, leading to enhanced splittings between otherwise degenerate or nearly degenerate orbitals and bands, reducing in many cases the kinetic energy as well. Bandwidths typically remain small, with overlap of the larger orbitals compensated by increase in the interatomic distances.
\begin{figure}
\centering
\includegraphics[width=0.9\linewidth]{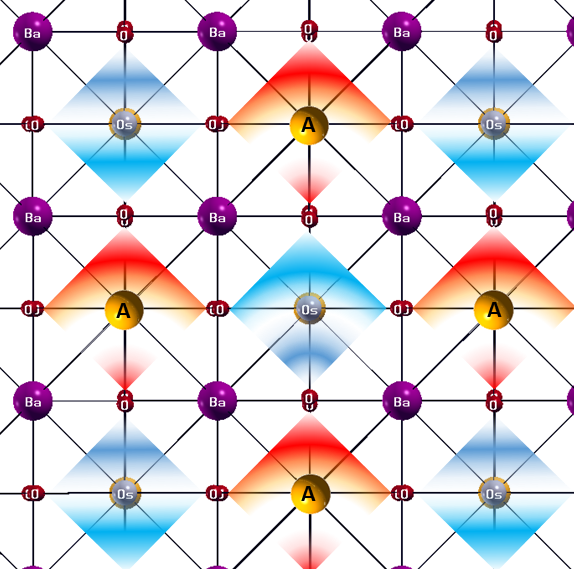}
\caption{Two dimensional schematic representation of face centered cubic double perovskites with a general formula Ba$_2$$A$OsO$_6$, blue and red shaded octahedra represent faces of Os-O $A$-O octahedra respectively ($A$=Na, Ca, Y). For these osmates the $A$ cation is closed shell and non-bonding, leaving each Os on with its own OsO$_6$ cluster.}
\label{fig1}
\end{figure}

In $5d$ $t_{2g}$ subshells where SOC remains unquenched (cubic symmetry), 
the six one-electron levels split into an upper 
$J$=$\frac{1}{2}$ doublet and a lower $J$=$\frac{3}{2}$ quartet. In this category Ir-based based magnets 
have been studied actively.\cite{ir1, ir2, ir3, ir4, ir5} A prominent class of such 
systems is the ordered double perovskites (DP), 
with chemical formula B$_{2}$$AA'$O$_{6}$. We are interested in the case where $A$ is a closed shell cation
and $A'$ is a magnetic ion; in such cases unusually high formal valence states can arise. A few examples 
attracting recent interest are
$A'$=Ru$^{5+}$ and Os$^{5+}$ in B$_{2}$Na$A'$O$_{6}$ (B=La and Nd),\cite{Nd2NaOsO6,la2nabo6prl, la2NaOs_ruO6} 
Mo$^{5+}$ in Ba$_{2}$YMoO$_{6}$, 
Os$^{6+}$ in Ba$_{2}$CaOsO$_{6},$ \cite{bcoo} 
and heptavalent Os in Ba$_{2}$$A$OsO$_{6}$ ($A$ =Li, Na). 
We chose three cubic Os based DP Ba$_{2}$NaOsO$_{6}$ (BNOO), Ba$_{2}$CaOsO$_{6}$ (BCOO) and Ba$_{2}$YOsO$_{6}$ (BYOO), where Os formal charge states varies across +7, +6, +5 with $d^1$, $d^2$ and $d^3$ formal configurations. 

High valent osmates can have an open $e_g$ shell rather than $t_{2g}$ shell.
Song {\it et al.}\cite{song2014} have reported a theoretical study\cite{song2014} of KOsO$_4$ with heptavalent Os, where large SOC,
strong correlations, and structural symmetry breaking conspire to produce an unexpectedly large
orbital moment in the $e_g^1$ shell that nominally supports no orbital moment. The moment arises due to distortion from ideal tetrahedral symmetry of  the ion, which allows mixing in of orbitals that support an orbital moment.

In this paper we present a comparative analysis of three cubic double perovskite osmates having $5d$ configurations, Ba$_2$$A$OsO$_6$, $A$=Na, Ca, Y. Experimental work on the Ca and Y analogues of BNOO show, as does BNOO, significant decrease of their magnetic moments from their spin only values compared to their $4d$ sister compounds. Probing this question is one focus of this paper.

 The experimentally studied compounds\cite{bnoobloo,fisherbnoo} 
Ba$_{2}$$A$OsO$_{6}$ 
show many features to make them of current interest. Besides their undistorted 
cubic double perovskite structure, 
and providing a sequence of heptavalent, hexavalent and pentavalent osmium 
compounds, BNOO is exotic in
being characterized as a {\it ferromagnetic Mott insulator},\cite{crystalbnoo,fisherbnoo} 
with order appearing at T$_C$ =6.8K with 
Curie-Weiss temperature $\Theta_{CW}=$ -10K. 
In BCOO, Curie-Weiss behaviour is observed for T$>$100 K 
with $\Theta_{CW}=$ −156.2(3) K. 
BCOO did not show any ordered moment in powdered neutron diffraction,\cite{bcoo1}, however from muon spin relaxation ($\mu$SR) data a long lived
oscillations exhibit a clear long-range magnetic order
below T$_C$ = 50 K. An estimate of the ordered moment on Os$^{+6}$ is around 0.2 $\mu$B, based
upon a comparison with $\mu$SR data for Ba$_2$YRuO$_6$ with a known ordered moment
of 2.2 $\mu$B.
The moment is much smaller than the spin-only value 
for a 5d$^2$ Os$^{+6}$ configuration, 
implicating a major influence of SOC.
BYOO exhibits a remarkably large Curie-Weiss temperature $\Theta_{CW}$=772 K indicating large
antiferromagnetic correlations within an fcc sublattice with high degree 
of magnetic frustration.\cite{byoo1} 
A magnetic transition to long range antiferromagnetic order, consistent with
fcc Type I order, occurs at $T_N$ = 69 K. The Os ordered moment of 1.65$\mu_B$ is 
much lower than the 
spin-only value of 3$\mu_B$.

Although BNOO's single $t_{2g}$ moment orders, the structure shows no evidence of 
the Jahn-Teller distortion that would accompany ordering of non-cubic charge density
that would destroy its cubic symmetry. The sister compound 
La$_{2}$NaOsO$_{6}$, on the other hand, with high-spin  $\mathit{d}^{3}$ Os configuration and a 
nominally cubic symmetry, is observed to be highly distorted.\cite{os_arrangment} This distortion 
can be understood in terms of geometrical misfit arising from incompatible ionic radii. 
There is another recent example of an Os-based based $5d^4$ perovskite compound 
BaOsO$_{3}$ that remains cubic;\cite{baoso3} 
on the other hand a related perovskite $5d^5$ NaOsO$_{3}$ does distort.\cite{naoso3} 
Again, this distortion may be due to differences in Goldschmidt's tolerance factor.

Lee and Pickett demonstrated in BNOO\cite{Pickett_bnoo} that, 
before considering magnetism and on-site interaction effects, SOC
splits the $t_{2g}$ bands into a lower $J$=$\frac{3}{2}$ quartet and an upper $J$=$\frac{1}{2}$ doublet,
as expected. Since BNOO is observed to be insulating and effects of spin-orbit coupling
drive the behavior, it provided the first
``J$_{\rm eff}$=$\frac{3}{2}$'' Mott insulator at quarter-filling, 
analogous to the ``J$_{\rm eff}$=$\frac{1}{2}$''
Mott insulators at half-filling that are being studied in $5d^5$ systems.\cite{ir4,jeff2,jeff3}
Since SOC is necessary to obtain the correct insulating ground state, 
these Dirac-Mott insulator states
provide a descriptive distinction to conventional $3d$ Mott insulators.

Including spin polarization and on-site Hubbard $U$ repulsion beyond the semilocal density
functional approximation within the DFT+U scheme\cite{AZA,Sasha,Erik} and including SOC, 
with both the WIEN2K and FPLO codes essentially full spin 
polarization was obtained but a gap was not opened\cite{Pickett_bnoo}
with a reasonable value of $U$.  
The basic complication seemed to be that the occupied orbital in
the OsO$_6$ cluster has only half of its charge on Os, with  the other half 
spread over the neighboring
O ions. Thinking in terms of hybridized local orbitals (viz. Wannier functions)
$U$ should be a value appropriate to the cluster orbital and 
should be applied to that orbital, however these codes apply $U$ only to the 
Os $5d$ orbitals. Xiang and Whangbo\cite{Whangbo2007} neglected
the Hund's rule $J_H$ in their DFT+$U$ calculation and did obtain a gap. 
However, neglecting $J_H$ omits both
the Hund's rule exchange energy and the anisotropy (orbital dependence) 
of the Hubbard interaction
$U_{mm'}$, whereas a full treatment should include all orbital dependencies to understand the
anisotropy on the Os site.

For this study we have applied a different orbital-specific hybrid DFT approach. 
The specific method, described in Sec. II,
gave a consistent explanation for BNOO of experimental observations\cite{fisherbnoo} of the 
(i) the Dirac-Mott-insulating state, (ii) the [110] easy axis, (iii) effective spin moment,
and (iv) the retention of the cubic structure. 
Here we apply this same approach for other two cubic Os-based DPs (BCOO and BYOO), 
having $d^2$ and $d^3$ Os respectively. All three DPs show a Dirac-Mott type behavior 
but SOC affects each somewhat differently. In Sec. III we outline briefly the theory
behind the NMR quantities that we report, in anticipation of such experiments. 
Results for band gaps and spin and orbital moments for three different directions of
magnetization are presented and discussed in Sec. IV. 
A radial $5d$ charge density analysis shows another revealing feature: 
though the formal charge of Os in these perovskites ranges from +7 to +5, the 
radial charge densities are virtually indistinguishable, {\it i.e.} the Os ions have the 
same physical $5d$ charge. This result is an extension of recent findings in other
transition metal oxides.\cite{yundi1,yundi2,yundi3}
The dependence of the nuclear magnetic resonance (NMR) chemical shielding factor
and susceptibility on the direction of spin is also presented.
Interpretation of NMR data in heavy $d$ oxides is still in an early stage;
one theoretical study of magnetic susceptibility and chemical shielding parameters 
for the $4d$ system Sr$_2$RuO$_4$ has published, along with formal details.\cite{pavarininmr}   
A short summary is provided in Sec. V.

\section{Computational Methods}
The present  DFT-based electronic structure calculations 
were performed using the full-potential augmented plane
wave plus local orbital method as implemented in the WIEN2K code.\cite{wien2k} 
The structural parameters of BNOO, BCOO and BYOO with full cubic symmetry of the DP
structure were taken from experimental X-ray
crystallographic data:\cite{crystalbnoo},\cite{bcoo},\cite{byoo1} $a$=8.28~\AA, $x_O$=0.2256; 
$a$=8.34~\AA, $x_O$=0.2410; $a$=8.35~\AA, $x_O$=0.2350, respectively. Note that the replacement
Ca$\rightarrow$Y does not change the lattice constant noticeably, but displaces the O ion somewhat. 
Atomic sphere radii, in a.u. were chosen as 2.50 (Ba), 2.00 (Na, Ca, Y), 1.80 (Os)
 and 1.58 (O). 
The Brillouin zone was sampled with a 
minimum of 400 k points during self-consistency, since coarser meshes were sometimes found
to be insufficient. 

\subsection{Onsite exact exchange: EECE functional}
For the exchange-correlation energy functional for treating strongly correlated insulators,
a variety of approaches in addition to DFT+$U$ exist and have
been tested and compared for a few selected systems.\cite{tran2006} 
In the DFT+$U$ approach, the relevant orbital is accepted to be, not an atomic orbital, but
rather that orbital hybridized with neighboring ligand orbitals, or (almost) equivalently, a
Wannier orbital. In the systems we treat this would be an OsO$_6$ octahedron cluster orbital.
However, most codes that implement the DFT+$U$ method use an atomic orbital (or
most of one, lying within an atomic sphere), and
apply the exchange energy shift $U$ and Hund's exchange shift $J_H$ to this orbital. 

However, the physical process that is not represented well in semilocal DFT is the
{\it exchange interaction on the transition metal ion} (here Os). More direct treatment of
exchange has been addressed with hybrid methods, which use for the
exchange functional a fraction of
local density exchange and another fraction (the complement of unity) of bare Hartree-Fock
exchange. Not only is this conventional hybrid approach computationally expensive, it
treats (and approximates) exchange interactions between itinerant states that may just as
well be treated by rapid semilocal methods. 

For this study we have chosen to apply the  
``exact exchange for correlated electrons''(EECE) functional
for treating correlated electron systems, introduced and evaluated by Novak and
collaborators.\cite{novak2006} Exact (Hartree-Fock) exchange is applied only to the
correlated orbitals ($5d$ for osmium) with full anisotropy of the exchange, 
and then implemented similarly to common use in hybrid
functionals, with 25\% of local density exchange being replaced by exact onsite exchange.
Although this EECE functional has not
been applied widely for Mott insulators, it has been found that it reproduces features
of the DFT+$U$ method -- opening of a gap in open-shell $d$ systems, for example --
without reference to any screened  Hubbard $U$ and Hund's rule $J_H$ parameters, 
and without facing the
choice of the type of double-counting correction\cite{Erik} to choose. As noted in the
Introduction, the EECE functional treatment makes the unusual Dirac-Mott 
insulating ferromagnetic ground state
of BNOO fully comprehensible.

We use this functional as implemented in the WIEN2K code,\cite{wien2kinput}
version 14.2. For the semilocal exchange-correlation part of the functional, the
parametrization of Perdew, Burke, and Ernzerhof\cite{PBE} (generalized gradient
approximation [GGA]) is used. SOC was included fully in core
states and for valence states was included in an accurate second-variational 
method using scalar relativistic wave functions,\cite{wien2ksoc}  a procedure that
is non-perturbative and quite accurate for $d$ orbitals even with large SOC.

This EECE method bears some kinship with to the more conventional 
hybrid exchange-correlation functionals (see
Tran {\it et al.}\cite{tran2006} for a comparison of several hybrid functionals). Hybrids replace
some fraction $\alpha$, typically 25\%, of local density exchange with Hartree-Fock
exchange, which then is approximated in various ways to reduce the expense to a more reasonable
level.  The EECE approach deals with exact exchange only for correlated orbitals,
however, making it appropriate for correlated materials. Note that it does not increase
bandgaps of ionic or covalent semiconductors, being more appropriate for local orbital systems. 

\section{NMR related quantities}
\subsection{Calculation of chemical shielding}
The chemical shift calculations are based on an all-electron linear response method where one obtains the induced current density considering the perturbation of the ground state wave functions due to the external magnetic field. The resulting magnetic shielding is then obtained by integration of the all-electron current according to the Biot-Savart law without further approximations.

Here, we outline the method of calculation of the NMR shielding. The approach 
is based on a linear response theory widely used for NMR calculations in 
solids.\cite{nmr1,nmr2,nmr3,nmr4} A detailed description of implementation of this approach 
into WIEN2K \cite{nmr1,nmr2,nmr3} was provided recently,\cite{nmr3,nmr4} 
and we summarize this approach here. 

The chemical shielding tensor $\overleftrightarrow{\sigma}$ is a proportionality factor 
between the induced magnetic field \textbf{$B_{ind}$} measured at the nucleus at site \textbf{R} and the external field \textbf{B}:
\begin{equation}
\textbf{B}_{ind}\textbf{(R)}=-\overleftrightarrow{\sigma}\textbf{(R)B}.
\end{equation}
For the materials we study, we deal with the isotropic shielding (IS), which is given by 
$\sigma\textbf{(R)} = Tr[\overleftrightarrow{\sigma}\textbf{(R)}].$ 
The experimentally measured chemical shift $\delta$ of the resonance line
is expressed with respect to some reference, so
$\delta\textbf{(R)} = \sigma_{ref} - \sigma\textbf{(R)}.$ 
The induced field $\textbf{B}_{ind}$ is evaluated by integrating the induced current $\textbf{j}_{ind}\textbf{(r)}$ according to the Biot-Savart law:
\begin{equation}
\textbf{B}_{ind}\textbf{(R)}= -\frac{1}{c}\int d^{3} r~\textbf{j}_{ind}\textbf{(r)} 
      \times \frac{\textbf{r}-\textbf{R}}{|\textbf{r}-\textbf{R}|^3}
\end{equation}
The calculations presented in this work have been performed using the 
WIEN2K\cite{nmr1} implementation
and are based on the augmented plane wave plus local orbital (APW+lo) 
method  with EECE\cite{novak2006} 
exchange correlation functional. Within this method, the unperturbed wave functions as well as
their first-order perturbations providing $\vec j_{ind}$
are expressed using plane waves in the interstitial region and an 
atomic-like representation inside
the atomic spheres. In contrast to standard ground-state calculations, evaluation of the isotropic
shielding requires an extended basis set inside the spheres, which is achieved by supplying 
additional local orbitals as described by Laskowski and Blaha.\cite{nmr1} To ensure convergence, 
we use ten extra local orbitals for Ba, the $A$ atom (Na,Ca,Y), Os, and O. 
For other computational parameters, 
the standard values lead to well-converged results.  
k point sampling of the Brillouin zone was done with an $7\times7\times7$ mesh in these 
insulating (or nearly insulating) states.

\subsection{Macroscopic magnetic susceptibility}
Application of a uniform magnetic field results in both spin and orbital magnetic responses, described
by the respective susceptibilities. The orbital magnetic susceptibility $\chi$ determines the 
macroscopic component of the induced orbital magnetic field, which may result in a contribution 
of several ppm to the chemical shielding.
To obtain the macroscopic susceptibility $\overleftrightarrow{\chi}$ that enters 
  $\textbf{B}_{ind} (\textbf{G}=0)$, 
we follow Mauri and Pickard \cite{pickard_mauri}, and use
\begin{equation}
\overleftrightarrow{\chi} = \lim_{q\rightarrow 0} \frac {\overleftrightarrow{F}(q)- 2\overleftrightarrow{F}(0) + \overleftrightarrow{F}(-q)}{q^2}
\end{equation}
where $F_{i,j} =(2-\delta_{i,j})Q_{i,j}$ and $i$ and $j$ are Cartesian indices. The tensor 
$\overleftrightarrow{Q}$ is given by 
\begin{equation}
\overleftrightarrow{Q}(q) = \Omega_{c} \sum_{\alpha=x,y,z}\sum_{o,\textbf{k}} 
  Re[\textbf{A}_{\textbf{k},q_{\alpha}}^o (\textbf{A}_{\textbf{k},q_{\alpha}}^o)^*]
\end{equation}
where $\textbf{A}_{\textbf{k},q_{\alpha}}^o$ are the matrix elements between Bloch states at 
\textbf{k} and \textbf{k} + \textbf{q}$_{\alpha}$

\begin{equation}
	\textbf{A}^{o}_{\textbf{k},\textbf{q}_\alpha} = \hat \alpha \times \left\langle u_{o,\textbf{k}}   \middle| 
	(\textbf{p}+ \textbf{k}) \middle| u^{(1)}_{\textbf{k}+\textbf{q}_{\alpha}} \right\rangle.
\end{equation}
$\left\langle u_{o,\textbf{k}}   \middle| (\textbf{p}+ \textbf{k}) \middle| u^{(1)}_{\textbf{k}+\textbf{q}_{\alpha}} \right\rangle$ are the matrix elements in the expression for the current density.
$\alpha$ is the Cartesian direction and $ u^{(1)}_{\textbf{k}+\textbf{q}_{\alpha}}$ 
is the periodic part of the perturbed wave function. 

\section{Electronic and magnetic characteristics }
An understanding of the effects of several competing energy scales in $5d$ oxides, viz. bandwidth,
repulsive interaction energy $U$, SOC strength, and magnetic exchange splitting, 
which in these systems are all of the
order of 0.5-1 eV, is only recently being explored in some detail. For this reason we provide
several results on this series of DPs to contribute to the building of understanding.  

\begin{figure}
  \includegraphics[width=0.9\linewidth]{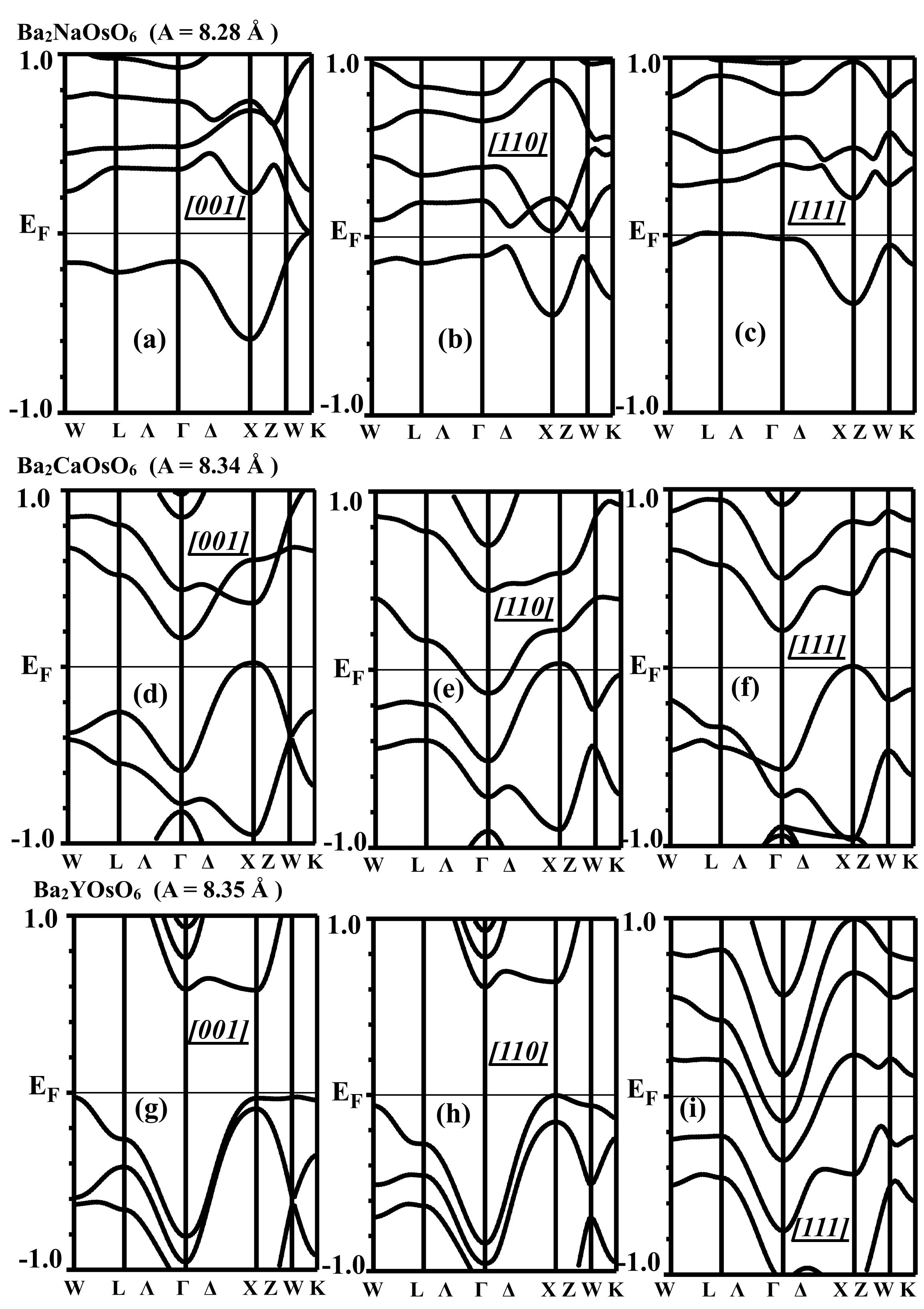}
\caption{Comparison band plots for Ba$_2$NaOsO$_6$ (a-c), Ba$_2$CaOsO$_6$ (d-f), Ba$_2$YOsO$_6$ (g-i)
	for the	three directions of magnetization axis (indicated on the plots).
Spin-orbit coupling coupled to magnetic exchange splitting leads to distinct W, X, and K points in
the zone. The chosen symmetry points for presentation
are, in units of $\frac{2\pi}{a}$: W=(1,$\frac{1}{2}$,0), L=($\frac{1}{2},-\frac{1}{2},-\frac{1}{2}$),
$\Gamma$=(0,0,0), X=(1,0,0), W=(1,$\frac{1}{2}$,0), and K=($\frac{3}{4},\frac{3}{4}$,0).}
\label{fig2}
\end{figure}

\subsection{Electron band dependence on spin direction}
It was shown previously that the band structure of BNOO shows strong dependence on the direction of the
spin, reflecting the strong SOC. The gap is small, and an insulator-metal transition (band overlap) occurs as the
direction of spin is rotated.  We show the corresponding behavior for BCOO and BYOO. 
We use experimental structures and ferromagnetic order to allow the most direct comparison, and
the progression with band filling in this series. 

Table I presents various contributions to the magnetism and the value of the calculated gap. 
Recall that the band filling increases $d^1$, $d^2$, $d^3$ in this sequence. 
The results are consistent
with expectations in most cases; orbital moments are small for the filled subshell 
$d^3$ case, although
by only a factor of 2-3 indicating some mixing of $e_g$ states. 
 The spin moment is closely  proportional to the band filling, as
expected for a fully polarized ion.
In all cases, the spin moment is reduced when SOC is included, by
5-15\%, and hardly depends on the direction of spin. The same is true of the orbital moment, with
one exception: for $d^2$ BCOO the orbital moment is much smaller for spin along [110] than for
the other two directions.

The occurrence, and value, of the gap is dependent on direction of spin (Table I), 
due to change of symmetry, as
shown earlier for BNOO. The relevant
part of the band structures are shown in 
Fig. \ref{fig2}, for each of the [001], [110], and [111] spin directions.
Fig. \ref{fig2}(a)-(c) corresponds to BNOO the different dispersion along chosen W-K directions. 
For spin along [001] [Fig. \ref{fig2}(f)] the valence band (VB) and conduction band (CB) almost 
touch. Upon decreasing the lattice parameter they overlap to give a metallic state, so
a insulator-metal transition under modest pressure is predicted. 
For other two spin directions such overlap does not occur at this volume.

Fig. \ref{fig2}(d)-(f) presents band plots for BCOO, calculated to be metallic only along [110]. 
Some of the larger changes can be seen along the W-K line. The important change is
the position of the lowest conduction band at the L point that is displayed; for [110]
spin direction a band overlap semimetallic state is obtained.
Fig. \ref{fig2}(g)-(i)) give the corresponding results for BYOO, strongly metallic for spin
along [111], where no incipient gap is identifiable.  
Whereas one might have expected SOC to have the largest effect for the open subshell
cases BNOO and BCOO, the effect is actually largest for BYOO, with a (relatively
speaking) large gap for two spin directions and considerable band overlap for
the third [111] spin direction. 

\begin{figure}
\centering
\includegraphics[width=0.9\linewidth]{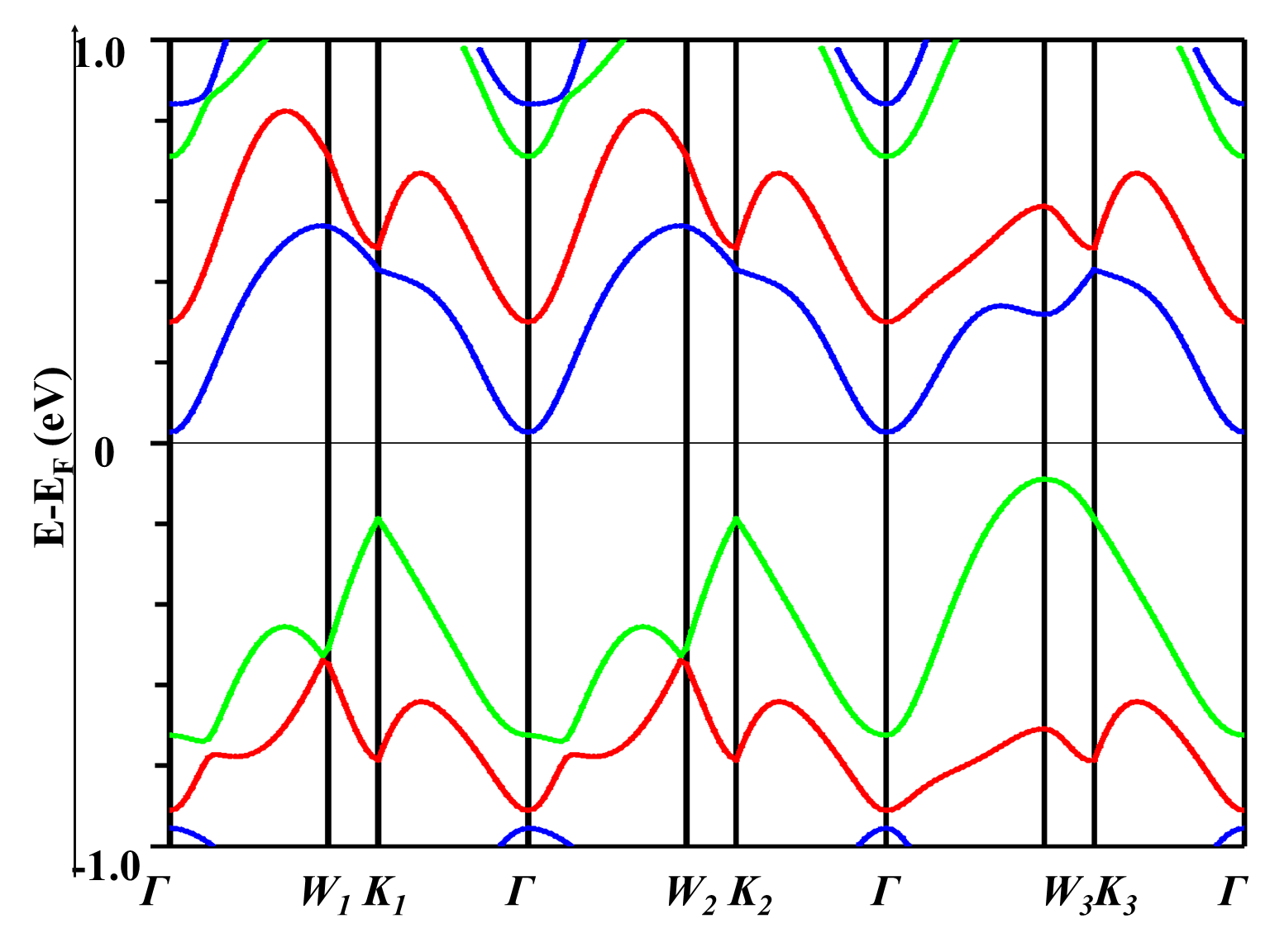}
\caption{Band plot along three different W-K directions for Ba$_2$CaOsO$_6$ with
	[001] magnetization axis ($a$= 8.35~\AA), illustrating the large effect 
    (up to 200 meV) of symmetry
	lowering by the large spin-orbit interaction strength. The chosen points, in units of
	$\frac{2\pi}{a}$, are $\Gamma$=(0, 0, 0),
	W$_1$=($-\frac{1}{2}$, 0 , 1),
	W$_2$=( 0 , -1, $\frac{1}{2}$),
	W$_3$=( 1 , 0 , -$\frac{1}{2}$),
	K$_1$=(-$\frac{3}{4}$, 0, $\frac{3}{4}$),
	K$_2$=( 0 , -$\frac{3}{4}$, $\frac{3}{4}$),
	K$_3$=($\frac{3}{4}$, 0, -$\frac{3}{4}$).}
\label{fig3}
\end{figure}

To be more explicit about the effects of symmetry lowering on the band structure 
by SOC, Fig. \ref{fig3} displays the bands for three (now inequivalent) W-K directions for
spin along [001] in BCOO, similar to that shown previously for BNOO.
Though the gap persists at a similar value along these lines, the linear band
behavior at the K point is very much different for the top valence band near
the right side of this figure, with interaction with the next lower band being
diminished.

\begin{table*}
    \begin{tabular}{c c c c c c c c c}
    \hline 
    \hline Method &  $\mu_{s}^{Os} $ &  $\mu_{orb}^{Os}$  & $\mu_{s}^{A}$  \ & 
           $\mu_{tot}^{Os} $ &  $\mu_{s}^{O_{x/y}}$  &  
           $\mu_{s} O_{z} $  &  Band Gap (eV) \\ 
    \hline
    \hline $Ba_2NaOsO_6$ $\mu_{interstitial}$=0.1 \\
    \hline
  EECE                  & 0.59 & N/A  & 0.0001  & N/A & 0.02  &  & No Gap \\
   EECE+SOC (001) & 0.52 & -0.41 &0.0001 & 0.11& 0.01 & 0.03 & 0.02 \\
   EECE+SOC (110) & 0.52  &-0.44  & 0.0001 & 0.08& 0.01  & 0.01 & 0.28  \\
   EECE+SOC (111)  & 0.52 & -0.45 & 0.0001& 0.07& 0.02  & 0.02 & 0.30 \\
    \hline $Ba_2CaOsO_6$ $\mu_{interstitial}$=0.3 \\
    \hline
    EECE & 1.32  & N/A  & 0.002  & N/A & 0.08 &  & No Gap \\
    EECE+SOC (001) & 1.14  &-0.61 & 0.002 & 0.53 & 0.09 & 0.04 & 0.13  \\
    EECE+SOC (110) & 1.17 & -0.24 & 0.002 & 0.98 & 0.08 & 0.04 & No Gap \\
   EECE+SOC (111) & 1.15  & -0.52 & 0.002 & 0.63 & 0.08 & 0.08 & 0.28  \\ 
    \hline $Ba_2YOsO_6$ $\mu_{interstitial}$=0.6  \\
    \hline
EECE & 1.8  & N/A & 0.02  & N/A & 0.09  & & No Gap \\
EECE+SOC (001) & 1.73 & -0.16 & -0.02 & 1.57  & 0.09  & 0.09 & 0.58 \\
EECE+SOC (110) & 1.72 &-0.11 & 0.01 &1.71 & 0.09 & 0.09 & 0.66  \\
EECE+SOC (111) & 1.74 & -0.13 & 0.02 & 1.61  & 0.09& 0.09 & No Gap \\
    \hline
    \end{tabular}
    \caption{Calculated spin, orbital, and total ($\mu_{tot}$=$\mu_s$+$\mu_{\ell}$) 
     moments ($\mu_B$) in the Os {\it sphere}, the spin moments on the O sites,
     and the band gap (eV) of Ba$_2$$A$OsO$_6$ ($A$=Na, Ca, Y). Values are presented
     for the three spin directions.
    The non-negligible interstitial contribution to the spin
    moment means that ``Os'' and ``O'' contributions are somewhat dependent 
    on the choice of sphere radii.}
\end{table*}

\subsection{Magnetocrystalline anisotropy}

BNOO has its experimental easy axis along the [110] direction. 
Our calculations successfully reproduced 
that result, however the energy for [111] is very close to that of [110]. The [001] spin
 direction sits substantially higher in energy. 
For ferromagnetic alignment, BCOO and BYOO both give
a [001] easy axis. The energy differences are material dependent, in the 1-10 meV range.
One feature in common for these three materials is that a metallic band structure
correlates with a higher energy, {\it i.e.} metallic phases are always disfavored.
This result is consistent with the observed insulating behavior observed 
in this series of compounds, and may result from extra kinetic energy of the conduction bands.

\subsection{Extent of SOC show spin up/dn matrix}
To illustrate the effect of SOC and spin direction in a $5d$ perovskite in this class, we tabulate in Table II the $5d$ spin-orbital occupations for $d^2$ BCOO for the three different directions. BCOO is  chosen because it has the largest orbital moments. Several noteworthy results can be discerned from Table II. First, though Os in BCOO is nominally fully spin-polarized $d^2$, the spin-orbital occupations indicate complexity behind this formal charge designation. The total $5d$ occupation number sum is around 3.4, not 2.0, moreover much of the $5d$ charge lies outside the sphere used to obtain these occupations. The largest spin-orbital occupation is 0.7; this should be understood as ``full occupation.'' Thus the actual $5d$ charge is actually 4.5-5, versus the formal value of 2. This
difference reflects the fact that formal charge is often quite different from real charge. 
This aspect of charge occupation is reasonably well recognized by electronic structure 
practitioners, and is important if one tries to understand the spin-orbital occupations. 
We return to the actual $5d$ charge issue in Sec. IV.E. 

Table II reveals, most broadly, that the net spin moment, and in fact its majority and minority contributions, are independent of spin direction, in line with the simplest picture of effects of SOC. Comparing the $m_l$ specific $5d$ occupation numbers, one notices that spin down occupations hardly change; there is minority spin occupation but in the net moment it remains out of the picture. The majority spin occupations do vary with spin direction, but in a more involved fashion than simple ionic models would predict. The two largest majority orbital occupations are emphasized in Table II. For each spin direction, there are two primary occupations(following the formal $5d^2$ charge state), but these vary with spin direction and the largest occupations differ in magnitude. For the various spin directions, the two ``occupied orbitals" are: [001], $m_l = -2, -1$; [110], $m_l = -2, +1$; [111], $m_l = -2, 0$. Only for [001] spin direction do these follow Hund's 2nd rule. These values reflect the strong environmental effects, that the localized orbitals are not simple Os $5d$ orbitals but rather are OsO$_6$ molecular orbitals where the orbital moment behavior becomes unpredictable without full calculation.   

\begin{table}
	\begin{tabular}{c c c c c c c c c}
		\hline \hline 
		SOC direction [001] & $m_{l}$ & -2 & -1 & 0 & +1 & +2 & Sum \\ 
		\hline
		Spin up & & {\bf 0.56} & {\bf 0.70} & 0.30 & 0.26 & 0.42 & 2.23  \\
		Spin down &    & 0.21  &      0.21  & 0.27 & 0.23 & 0.26 & 1.19  \\
		Net spin & & {\bf 0.35}  &  {\bf 0.49}  & 0.03 & 0.03 & 0.16 & 1.04\\
		\hline
		SOC direction [110] & $m_{l}$ & -2 & -1 & 0 & +1 & +2 & Sum \\ 
		\hline
		Spin up & & 0.40 & {\bf 0.58}  & 0.29 & {\bf 0.58} & 0.40 & 2.24  \\
		Spin down & & 0.24 & 0.21      & 0.27 &      0.21 & 0.24 & 1.18  \\
        Net spin & &      0.16  &{\bf 0.37}  & 0.02 & {\bf 0.37} & 0.16 & 1.06\\
		\hline
		SOC direction [111] & $m_{l}$ & -2 & -1 & 0 & +1 & +2 & Sum \\ 
		\hline
		Spin up & & {\bf 0.61} & 0.31  & {\bf 0.69} & 0.39 & 0.25 & 2.24  \\
		Spin down & &    0.21 &  0.24 &       0.20 &  0.27 & 0.26 & 1.18  \\
        Net spin & &  {\bf 0.40}  &      0.07  &{\bf 0.49} & 0.12 &-0.01 & 1.06\\
		\hline \hline
	\end{tabular}
	\caption{5\textit{d} orbital specific occupation for Ba$_2$CaOsO$_6$, 
  using three spin directions [001], [110], [111]. Spin up and spin down 
  orbital occupations along with the difference (Net spin) are displayed;
  the meaningful ones are in boldface and appear clearly in the net spin. 
  The total occupation sum of the \textit{5d} shell (right hand column)
  remains the same for the
  different spin directions.}
\end{table}

\begin{figure}
    \centering
    \includegraphics[width=0.9\linewidth]{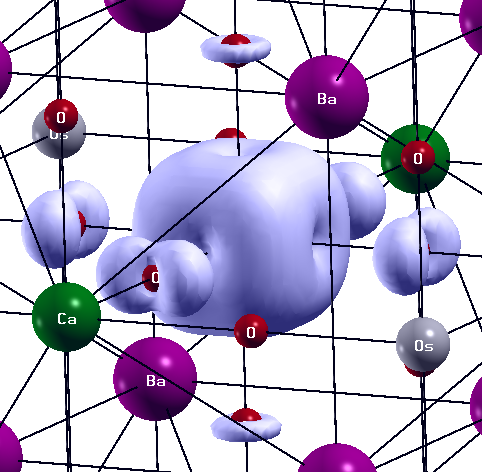}
\caption{Isosurface plot of the spin density (equal to the valence charge density)
  of $d^2$ BCOO for [001] spin
  orientation. The near-cubic shape of the density on the Os ion
  supports the lack of any discernible Jahn-Teller distortion from cubic symmetry. 
 There is more induced spin density on the equatorial O ions than on the apical ions.
  This plot was generated by XCRYSDEN.\cite{xcrysden} }
    \label{fig4}
\end{figure}

\subsection{Spin and orbital moments}
The EECE functional was applied previously\cite{bnoo_prb} to understand observed behavior 
in BNOO, providing a description of the ferromagnetic Mott insulating ground state, the 
[110] easy axis, and the lack of Jahn-Teller distortion which leaves BNOO with the 
undistorted cubic structure.  We apply this same protocol to BCOO and BYOO, to assess its 
performance and to understand these two materials better. Although BCOO and BYOO, unlike  
BNOO, order antiferromagnetically,\cite{byoo1,bcoo1} we make comparisons using ferromagnetic 
alignment of moments. (The magnetic ordering question becomes very involved when SOC is 
large, resulting in exchange coupling between mixed spin-orbitals.\cite{SYS1,SYS2}) 
As in the previous work on 
BNOO,\cite{bnoo_prb} calculations are first converged without SOC, then the direction of 
spin is chosen and SOC is applied. Several results are presented in Table I. The spin 
moment of Os ($\mu_s^{Os}$) increases linearly with formal charge of Os. $\mu_s^{Os}$ 
values are decreased somewhat by SOC, more so for BCOO. 

The orbital moment always opposes the spin moment, in agreement with Hund's third rule.
For $d^1$ BNOO and $d^3$ BYOO $\mu_{orb}^{Os}$ are almost invariant with respect to spin direction. BCOO behaves quite differently: $\mu_{orb}^{Os}$ is -0.6 to -0.5$\mu_B$ for [001] and [111] directions, but is somewhat less than half that for spin along [110]. Thus the orbital moment of Os on BCOO is larger, as expected, than on the other two materials, but also is sensitive to the direction of the spin, an unappreciated aspect of magnetocrystalline anisotropy.
The contributions of the various O ions differ depending on the symmetry remaining after SOC is included. 

The observed low temperature ordered moment on BYOO is 1.65$\mu_B$,\cite{byoo1} which is 
much reduced from the expected spin only value of 3$\mu_B$ (and no orbital contribution 
from the filled subshell). Table 1 gives the calculated Os moment {\it within its sphere} 
in  BYOO of 1.57--1.71 $\mu_B$, depending on the direction of spin. Our magnetocrystalline 
anisotropy energy calculation shows [001] is the easy axis for ferromagnetic ordered BYOO; 
in that case the Os effective magnetic moment is 1.57 $\mu_B$, naively close to the 
experimental value.\cite{byoo1} 
However, the net moment should be equal to the {\it total} spin moment minus the 
{\it total orbital} moment. The full spin moment (of the cell, {\it i.e.} of the OsO$_6$ 
cluster) is 3$\mu_B$. From the orbital occupations of Table II, the largest value is 0.70, 
meaning that the remaining 30\% lies outside the sphere, so the sphere orbital moments 
should be scaled up by $\frac{1}{0.7}=1.43$. For Os in BYOO, this is about 0.2$\mu_B$, making the 
net moment 2.8$\mu_B$,  considerably larger than the observed ordered moment. The resolution 
to this paradox is that BYOO is actually AFM ordered, in which case the spin moment of the 
OsO$_6$ cluster may be, and no doubt is, reduced from the full 3$\mu_B$ value. Thus our 
FM moment provides no means to compare with the observed AFM value. 

Above it has been suggested that the spin moment in the FM state is reduced
compared to our calculated (FM) moment, which based on experience is likely
though the amount of reduction is uncertain. In any case, the question of the
BNOO moment remains, as does the issue of FM order in BNOO, whereas the sister
compounds BCOO, BYOO, and Ba$_2$LiOsO$_6$ all show AFM order. A possible resolution
is that the FM order in BNOO is actually canted AFM order that is, the basic
exchange coupling is antiferromagnetic between nearest neighbors, but this is
frustrated in the fcc sublattice and the balance it tipped by second neighbor
coupling or differences in the character of the spin-orbital that is occupied.
With canted AFM order, the ordered moment is some fraction of the net local
moment. Neutron scattering data might resolve these questions. 

\begin{figure}
\centering
\includegraphics[width=0.9\linewidth]{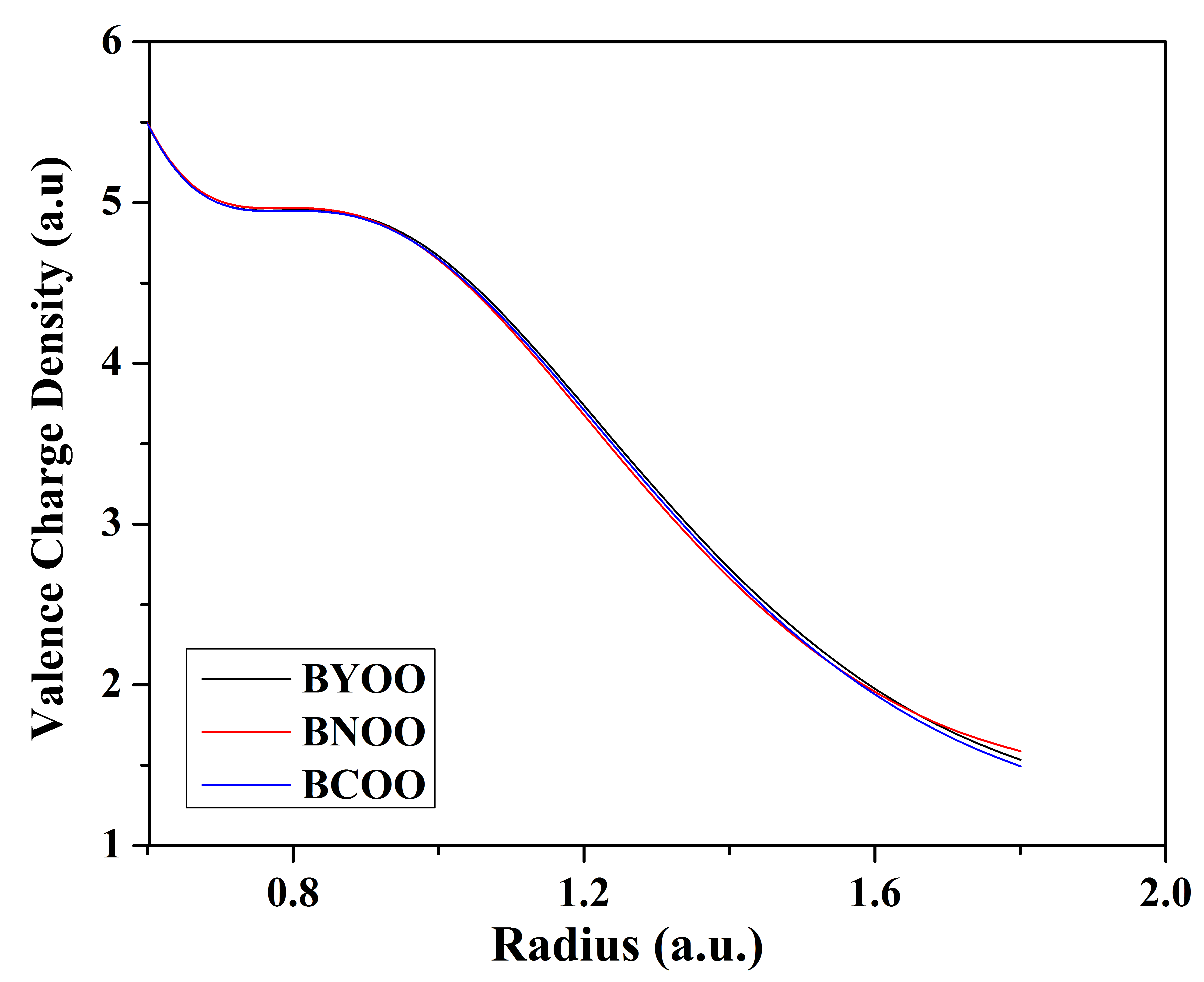}
\caption{Plot of the radial valence charge density on Os for the three 
  double perovskites, restricted to the region outside the $4f$ shell.
  The $5d$ density peak occurs around 0.9 a.u.
The three curves coincide until near the sphere boundary, where the
charge is affected by tails from neighboring O ions. There is
negligible difference in the $5d$
peak region, indicating no difference in $5d$ charge on Os for this sequence with formal
charges of $d^1, d^2,$ and $d^3$. }
\label{fig5}
\end{figure}

\subsection{Radial charge density analysis}
This sequence of three descending formal valence osmates provides the opportunity to address recent developments in the understanding of formal charge states.\cite{yundi1,yundi2,yundi3} The $d^1, d^2, d^3$ sequence suggests substantially differing charges on the Os ions, even if the unit difference in formal charge is not expected. Several cases have arisen recently where charges on differing formal charge states have produced perplexing results: the actual charges don't differ.
Fig. \ref{fig5} presents the radial charge densities for all three double perovskites in this series. Notwithstanding the formal charges +7, +6, +5, or configurations $d^1, d^2, d^3$ respectively, the radial charge density in the region of the peak in $5d$ charge shows no difference, down to the percent level.
This is surprising but consistent with several other examples; it is simply too costly energetically for ions to have differing charges. The formal valence concept remains an important concept, but is a property of an ion and its environment rather than of the ion alone.

This constancy of $5d$ charge on Os prompted us to look at the sphere charges of the other atoms. The Ba and O atoms, which are common to this sequence of compounds, have nearly identical radial charge densities across the series (as their formal valences suggest), with roughly +3 and -1.3 net charges within the spheres used in the calculations. Note that these do not and should not closely reflect the formal charges, partly because inscribed (non-overlapping) spheres are used, and much charge remains in the interstitial region.  
Na, Ca, and Y sphere charges  should not be compared, as the same sphere size has been used for ions of differing charge.

\begin{figure}
    \centering
    \includegraphics[width=0.9\linewidth]{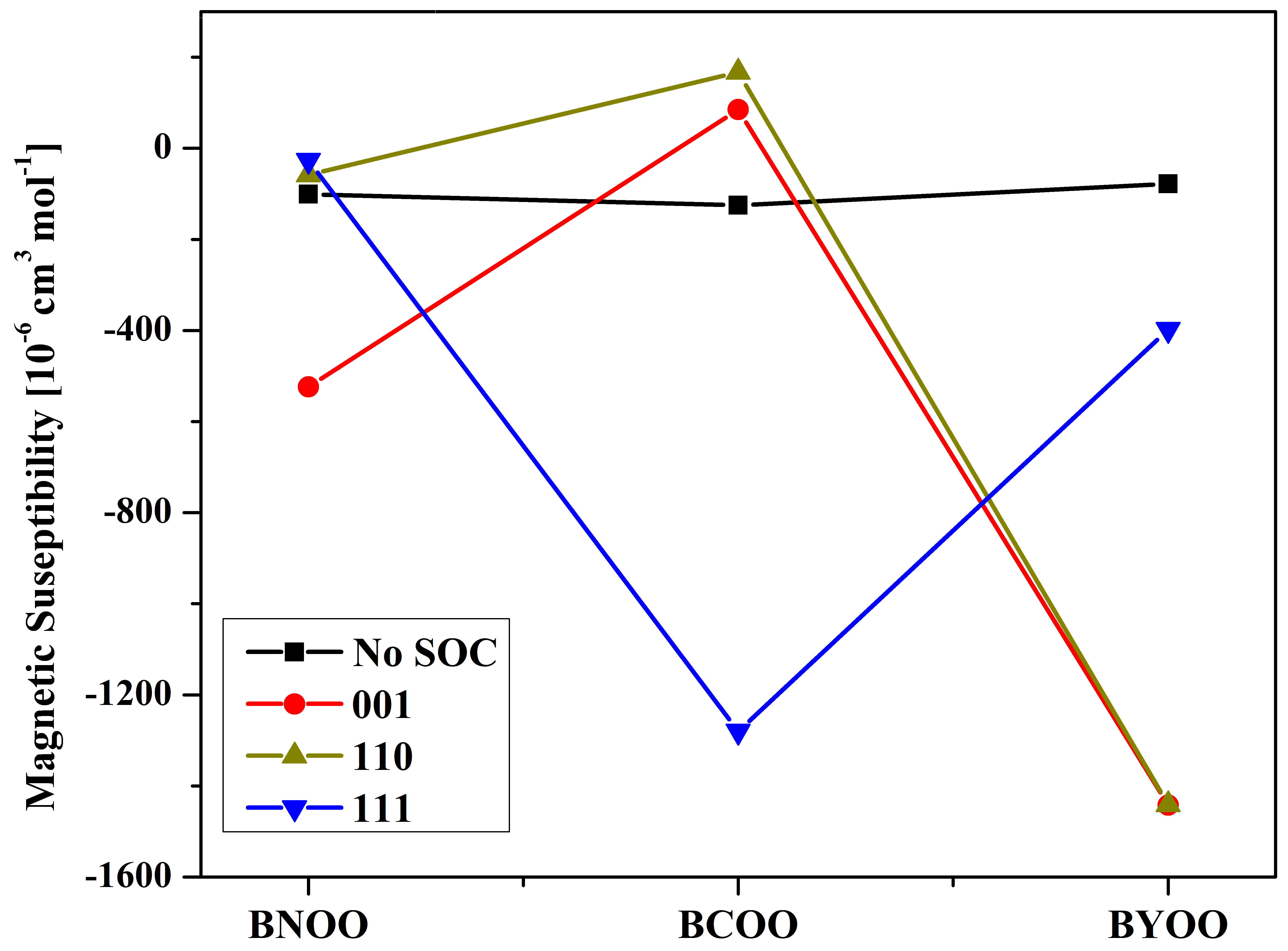}
    \caption{Macroscopic magnetic susceptibility $\chi$ calculated for BNOO,
BCOO, and BYOO, first without spin orbit coupling, then for the three different
spin directions [001], [110], [111]. Without spin orbit coupling $\chi$ is the
same (and small) for all three.  Including spin-orbit coupling results in 
substantial changes, non-monotonic behavior across the series, and dependence
on spin orientation.}
    \label{fig6}
\end{figure}

\subsection{Magnetic susceptibility and chemical shielding } 
There is little intuitive understanding for the types of compounds
being studied here of what characteristics, and what
energy ranges, of wavefunctions influence
the  macroscopic magnetic susceptibility $\chi$ and the chemical shielding 
fraction measured in NMR experiments. It is unclear for example if states
near the band edge are particularly important, and just how large (strong)
spin-orbit coupling effects are. We  have evaluated these properties
as described in Sec. III.B.

For consistency in the comparison, we have used 
the same muffin tin radii for Ba, Os, O for all three compounds, and for 
Na, Ca, Y also same values of muffin tin radius were used.

Fig. \ref{fig6} shows the macroscopic orbital susceptibility $\chi$ values for
the three compounds, without SOC and then including
SOC to assess the effect. Without SOC all are predicted to have
 almost the same small diamagnetic value of $\chi$. SOC results in
substantial changes: some of the values become strongly diamagnetic,
and the behavior in non-monotonic with band filling.
For [001] and [110] spin directions $\chi$ follows
a similar trend, a positive trend from $d^1$ to $d^2$ ({\it i.e.} less
negative) followed by a steep diamagnetic swing for $d^3$ BYOO. Another
way to look at these results is that $d^2$ BCOO contains more paramagnetic contributions
to $\chi$ than do the other two. The prediction is that the spin direction
can be as important in determining $\chi$ as the band filling, and that
there is no clear correlation (with spin direction, with filling, or with
bandgap) in this small sample. In short, SOC is the determining factor in the
susceptibility of these compounds.

\begin{figure}
\centering
\includegraphics[width=1.0\linewidth]{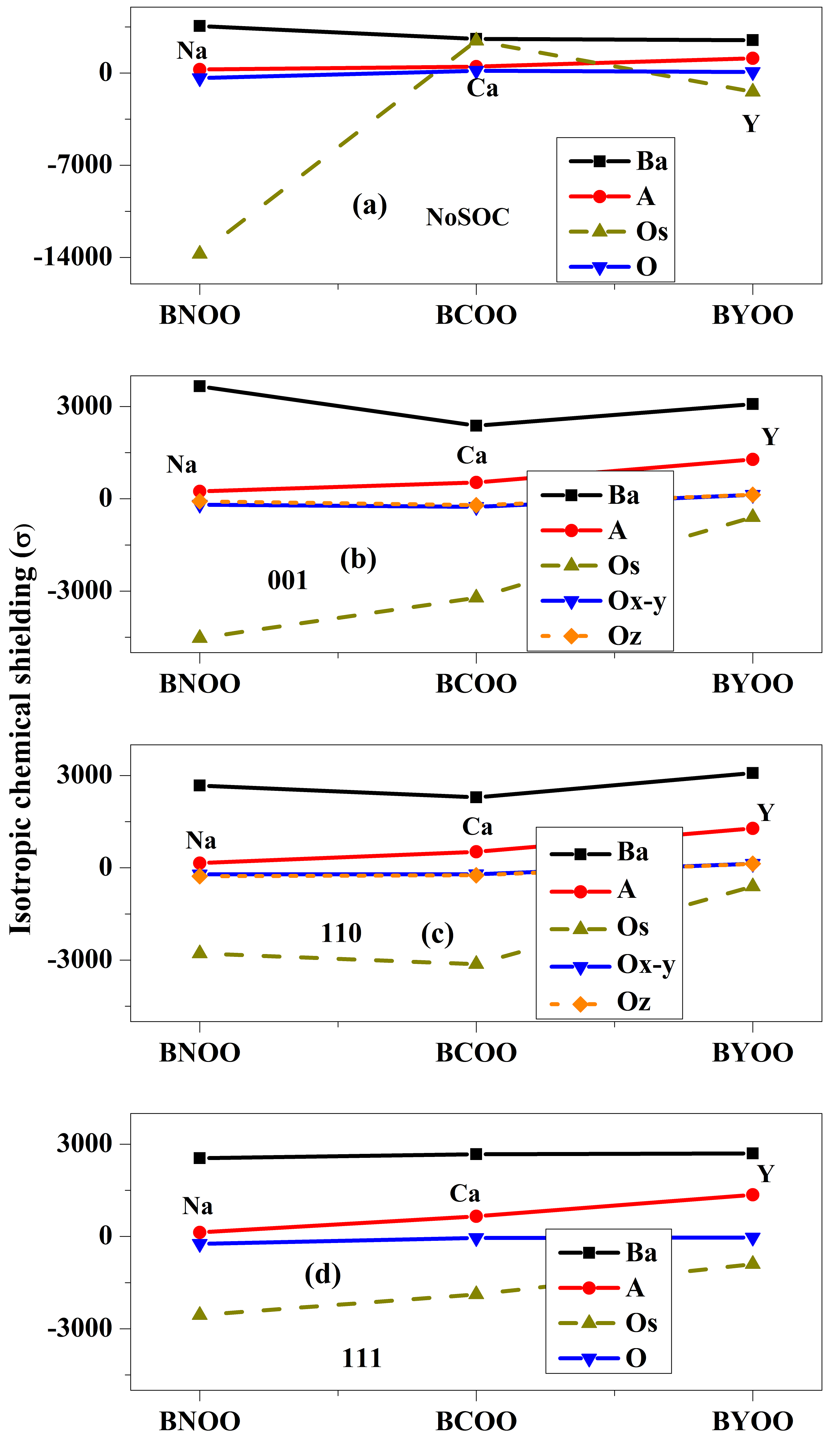}
\caption{Isotropic shielding fractions $\sigma$ (in ppm) calculated  as
described in the text.
Results are shown (a) without, and (b)-(d) with, spin orbit coupling 
along three spin orientations: 
(b) [001], (c) [110] (d) [111]. } 
\label{fig7}
\end{figure}

Fig. \ref{fig7}(a) presents the isotropic chemical shielding parameters before
inclusion of spin-orbit coupling. 
Except perhaps for Os, the values of $\sigma$  for
Ba, O and the closed shell $A$ cations vary little across the three compounds. 
The values for Na and O
are very small, for Ba are positive and larger (3000 ppm). The Os value has
more interesting non-monotonic behavior, changing from 
large and negative for BNOO to small and positive for BCOO, then small and negative for BYOO. 
Evidently the polarizable $5d$ electrons in a $d^1$ configuration
provide a better platform for induced orbital currents.

Spin-orbit coupling alters the chemical shielding fraction behavior, especially for
Os.  For all three directions
of spin Fig. \ref{fig7}(b)-(d), 
$\sigma$ for Os becomes monotonic with formal valence and not seriously 
deviating from linear while staying negative.
SOC makes only small changes in $\sigma$ for Na and O ions, staying
about the same for all three compounds. 
When there are two O sites due to SOC lowering of symmetry, Fig. \ref{fig7} (b)-(c), 
there is no difference in $\sigma$. For Ba, $\sigma$ increases substantially to
2500-3000 ppm and is relatively constant across the series.
Experimental data will be extremely useful in understanding what can be 
understood and extracted from
the calculated values.

Pavarini and Mazin\cite{pavarininmr} suggested that NMR related properties are mainly tuned 
by the orbital participation near Fermi level in the correlated metal Sr$_2$RuO$_4$ that
they studied, which is known to harbor spin fluctuations.
These osmates are insulating for two of the three spin directions, and the
shielding fractions do not seem to be influenced much by the small metallic
density (when it is present).  
Fig. \ref{fig2} indicates that bands near the gap have only significant Os $5d$ 
and O $2p$ character.  Fig. \ref{fig7} (a) shows Os that, before including SOC,
the shielding fraction on Os does not follow linear behavior versus formal
$d$ occupations. However, when SOC is applied, Fig. \ref{fig7}(b)-(d),
monotonic (or nearly so) progression is restored. Not surprisingly, the
shielding fraction on the other ions is hardly affected by SOC.

\section{Summary}
The EECE density functional, in which exact exchange in the Os $5d$ shell
is combined in hybrid fashion with local density exchange,
has been used to model the sequence of increasing  formal $5d^n$ 
osmates Ba$_2$$A$OsO$_6$, $A$ = Na, Ca, and Y.   
Spin-orbit coupling is essential in opening the gap, putting all three
in the ``Dirac-Mott'' class of correlated insulators. The gaps are
predicted to be small (a few tenths of eV) and rotation of the direction
of magnetization can produce an insulator-metal transition through band
overlap. For insight into the spin and orbital nature of $5d$ ions, we 
presented an example of the effect of SOC through the \textit{5d} 
orbital occupation matrix. Irrespective of SOC direction the total 
occupation stays constant, but orbital specific occupation depend
considerably on direction of magnetization, leading to different 
orbital moments. 

For comparison with (near) future NMR data we have
presented the effect of SOC on the macroscopic susceptibility and the 
chemical shielding fraction.  Spin-orbit coupling has a substantial effect
on the susceptibility as well as the shielding fraction.
Comparison of the radial charge densities
reveals that the three differing formal charge states have the same 
physical Os $5d$ charge, a surprising result but one consistent with
recent finding in $3d$ ``charge ordered'' systems. This result is
more striking when it is realized that formal charge states differing
by two ($d^1$ versus $d^3$) have the same physical charge. This result
re-emphasizes that formal charge is a property of an ion and its
environment, which is not well recognized.

\section{Acknowledgments}
We acknowledge many useful conversations with K.-W. Lee and comments on the
calculations from A. S. Botana.
Our research used resources of the National Energy Research Scientific Computing 
Center (NERSC), a DOE Office of Science User Facility supported by the Office of 
Science of the U.S. Department of Energy under Contract No. DE-AC02-05CH11231.
S.G. was supported by U.S. DOE Stockpile Stewardship Academic Alliance
Program under Grant DE-FG03-03NA00071, and W.E.P. was supported by 
U.S. DOE Grant DE-FG02-04ER46111.


\begin{thebibliography}{27}
\bibitem{wien2kinput}In the WIEN2K {\rm case.inorb} file the flag {\rm HYBR} is used
for Os $5d$ orbitals, with a fraction of 0.25.
\end{thebibliography}
\end{document}